\documentclass[twocolumn,aps,prl,groupedaddress]{revtex4}
\usepackage{epsfig,amssymb,amsmath}
\usepackage[hypertex,linkcolor=red]{hyperref}

\usepackage{color}

\def\comment#1{}
\def\labell#1{\label{#1}}
\def\section#1{{\par\em #1:--- }}
\def\togli#1{}

\def\>{\rangle}
\def\<{\langle}
\begin{document}

\title{ Efficient universal blind computation} \author{Vittorio
  Giovannetti$^1$, Lorenzo Maccone$^2$, Tomoyuki Morimae$^{3,4}$,
  Terry G.  Rudolph$^3$} \affiliation{ \vbox{ $^1$NEST, Scuola Normale
    Superiore and Istituto
    Nanoscienze-CNR, piazza dei Cavalieri 7, I-56126 Pisa, Italy}\\
  \vbox{$^2$Dip.~Fisica ``A.~Volta'', Univ.~of Pavia, via Bassi 6,
    I-27100 Pavia, Italy} \vbox{$^3$Department of Physics, Imperial
    College London, SW7 2AZ, United Kingdom} \vbox{$^4$ASRLD Unit,
    Gunma University, 1-5-1 Tenjin-cho, Kiryu-shi, Gunma 376-0052,
    Japan} }
\begin{abstract}
  We give a cheat sensitive protocol for blind universal quantum
  computation that is efficient in terms of computational and
  communication resources: it allows one party to perform an arbitrary
  computation on a second party's quantum computer without revealing
  either which computation is performed, or its input and output.  The
  first party's computational capabilities can be extremely limited:
  she must only be able to create and measure single-qubit
  superposition states. The second party is not required to use
  measurement-based quantum computation. The protocol requires the
  (optimal) exchange of $O(J\log_2(N))$ single-qubit states, where $J$
  is the computational depth and $N$ is the number of qubits needed
  for the computation.
\end{abstract}
\maketitle

Blind computation allows one party (say Alice) who has limited
computational power, to use the computational resources of another
party (say Bob), without revealing which computation she performs, nor
her input and output data. As one expects, arbitrary blind computation
is impossible using a classical computer \cite{abadi}. Surprisingly,
arbitrary blind computation is instead possible on a quantum computer
\cite{childs,bfk,Barz,Vedran,AKLT,FK,MF_topo,MF_MA,CV,NS,Sueki},
achieving unconditional security premised only on the correctness of
quantum physics, similar to that achieved for quantum key
distribution.  Here we propose a blind universal computation protocol
that is efficient in terms of communication between Alice and Bob.
Differently from previous
proposals~\cite{childs,bfk,Barz,Vedran,AKLT,FK,MF_topo,MF_MA,CV,NS,Sueki}
our scheme is based on a cheat-sensitive strategy: Alice can detect
whether a dishonest Bob is trying to ascertain the computation she
wishes to perform.  Moreover, in contrast to the one-time-pad protocol
of Childs \cite{childs}, ours does not require the computational
qubits to be exchanged between Alice and Bob and, in contrast to the
BFK protocol \cite{bfk}, it does not require measurement-based
computation, but is described in the circuit model.  Our protocol
requires only $O(J\log_2 N)$ qubits to be exchanged between Alice and
Bob ($J$ the computation depth, $N$ the number of qubits required for
the computation), achieving an exponential gain in $N$ in
communication complexity over previous protocols, that require $O(NJ)$
communication overhead.  As in the previous schemes, Alice's
computational capabilities can be extremely limited: she must only
generate and measure single-qubit states in the computational
$\{|0\>,|1\>\}$ or the complementary
$\{|\pm\>=(|0\>\pm|1\rangle)/\sqrt{2}\}$ bases.

The main idea is that Alice communicates to Bob the gates that he
needs to apply using a Quantum Private Query (QPQ)-inspired protocol
\cite{qpq,qpq2}: she encodes this information into a quantum register
and she randomly intersperses her communication with decoys
\cite{arrighi}. Bob must apply the gates blindly and send back her
register without extracting information from it. If he does try to
extract the information, Alice can detect this from a single-qubit
measurement of the decoy states she received back and she can
interrupt her computation. In this way Bob can determine at most a
constant number of steps of the computation before Alice has a high
probability of detecting that he his cheating.  Asymptotically in $J$,
he thus obtains no information on her computation. Moreover, it is
easy for Alice to hide both her input data (since the encoding of the
input state is part of the computation) and her output data (since she
can instruct Bob to make random flips of the output state bits prior
to the final measurement).  Finally, by adapting the approach proposed
in \cite{abe,bfk,FK,NS}, our scheme can allow the
computationally-limited Alice to test whether Bob is performing the
computation requested.

\section{The protocol} Bob controls a quantum computing facility which
includes a quantum memory ${\cal M}$ composed of $N$ qubits
initialized in a fiducial state (say the vector $|0\rangle^{\otimes
  N}$) and a set ${\cal G}$ of $O(\mbox{poly}(N))$ universal gates he
can apply to them. To fix the notation, we can assume for instance
that ~${\cal G}$ contains ${G}=N(N+2)$ elements including a Hadamard
and a $\pi/8$ gate for each qubit and a C-NOT for each (ordered)
couple of qubits of ${\cal M}$~\cite{univ}. In this scenario, Alice
can instruct him to perform an arbitrary computation by telling him
which element of ${\cal G}$ he must apply at each step of the
computation. For example, Alice can send a number $n$ between $0$ and
${G}-1$, which Bob interprets in the following way: $0$ to $N-1$ means
``act with a Hadamard gate on qubit $n$'', $N$ to $2N-1$ means ``act
with a $\pi/8$ gate on qubit $n-N$'', and any other number means ``act
with a C-NOT gate using $n_1$ as control and $n_2$ as target'', where
$n_1,n_2$ are such that $n=n_1N+n_2+2N$. For this code, she needs a
$\log_2 {G} \simeq O(\log_2 N)$ bit register.  She can thus instruct
Bob to perform an arbitrary $J$-step computation by giving him $J
\log_2 {G}\simeq O( J \log_2 N)$ bits.  This communication cost is
optimal, because a programmable quantum computer requires a program
register of dimension at least as large as the number of possible
computations that it can perform \cite{nc} (since at each step Bob can
apply one out of ${G}$ possible gates, in our case such number is
indeed equal to ${G}^J$, which requires $J\log_2G$ bits).

To achieve blind computation, Alice must intersperse her instructions
to Bob with decoy queries (we shall see in the following that this can
be done with a linear overhead in terms of communication and size of
the memory ${\cal M}$). Namely, at each computational step, she sends
a register $A$ of $\log_2{G} \simeq O(\log_2N)$ qubits.  This register
contains either a {\em plain instruction} for Bob or, at random times,
a {\em quantum decoy}.  Plain instructions are encoded by preparing
$A$ in a state $|n\rangle_A$ of the computational basis: for $n\in\{
0, \cdots , {G-1}\}$, it indicates to Bob that the $n$-th element
$U_n$ of the set ${\cal G}$ must be applied to the memory ${\cal M}$.
In contrast, quantum decoys are prepared by creating superposition of
instructions of the form $\sum_{n\in D} \eta_n |n\>_A$ where $D$ is a
subset of $\{0, \cdots , {G-1}\}$ containing at least two elements,
$\eta_n \in \{ -1,1\}$ (for ease of notation, normalizations are
dropped). To do so, as in the BB84 protocol~\cite{bb84}, it is
sufficient for Alice to initialize her register $A$ into factorized
states where some of the qubits are in the computational basis
$\{|0\>,|1\>\}$ while the others are in the basis $\{|+\>,|-\>\}$ (a
task she can achieve even with limited computational capabilities).
For instance, (in the simplified case of two-qubit register) she can
produce a quantum decoy where the instructions $|n=0\rangle_A$ and
$|n=2\rangle_A$ are superimposed by sending the qubits of $A$ in the
factorized state $|+,0\rangle_{A} = |0,0\rangle_A + |1,0\rangle_A$
\cite{NOTA1}.

Since Alice is sending plain messages and decoys at random, Bob cannot
perform even partial measurements in the computational basis without
risking disrupting the coherence of Alice's superpositions. Still he
can use the register $A$ (without measuring it) as a quantum control
to blindly trigger his operation on ${\cal M}$ e.g. by employing a
qRAM (quantum Random Access Memory)~\cite{qram,ncl}.  Specifically,
for each state $|\Psi\rangle_{\cal M}$ of ${\cal M}$ and
$|\phi\rangle_A = \sum_n\alpha_n |n\rangle_A$ of $A$ he performs the
control-unitary gate $U_{Bob}=\sum_n|n\>\<n|\otimes U_n$ which yields
the mapping
\begin{eqnarray}\label{ram} |\phi\rangle_A
      \otimes |\Psi\rangle_{\cal M}\;\;\;
{\longrightarrow}\;\;\; 
\sum_n \alpha_n |n\rangle_A\otimes U_n |\Psi\rangle_{\cal M}\;.
\end{eqnarray}
After this operation, Bob sends the register $A$ back to Alice. Alice
now checks her decoys to verify that Bob, when processing $A$, did not
try to ``read'' its information content. There are two possible cases:
her decoys can be unentangled from Bob's qubits or they can be
entangled. 

Case (a): The first case happens whenever Alice's decoy is a
superposition of $n$, $n'$, $n''$,..., such that
\begin{eqnarray} \label{constraint} U_{n} | \Psi\rangle_{\cal M} =
  U_{n'} |\Psi\rangle_{\cal M}=  U_{n''} |\Psi\rangle_{\cal M}=... \;,
\end{eqnarray} 
(e.g.  Alice, through the register $|n\>_A + |n'\>_A$, instructs Bob
to apply a $\pi/8$ gate to a superposition of two qubits of ${\cal M}$
which she knows are initially in $|0\rangle$). In this case, if Bob
follows the protocol, the final state of $A$ and ${\cal M}$ is
factorized and the $A$ register is unchanged,
see 
Eq.~(\ref{ram}).  In contrast, if Bob had measured the register $A$
(or had entangled it with an ancilla), the superposition will be
corrupted.  Alice can exploit this to monitor Bob: she measures the
register he sent back (single-qubit measurements suffice) and checks
that the results match the qubits she had sent Bob. If they do not
match, the superposition of one of her decoys was collapsed: she knows
that Bob is trying to find out the values encoded in her registers and
she stops the computation.  Otherwise, she can assume that Bob has not
obtained information on which computation she instructed him to
perform. Alice can easily enforce Eq.~\eqref{constraint} by devoting a
(random, secret) subset of Bob's memory $\cal M$ to doing trivial
operations on decoy states (e.g.~keeping some qubits in that subset in
a state $|0\>$ on which decoys composed of $\pi/8$ gates and C-NOTs
act trivially, and other qubits in eigenstates of the Hadamard, on
which Hadamard decoys act trivially).

Case (b): The second case happens whenever Eq.~\eqref{constraint} is
not satisfied and the register $A$ becomes entangled with Bob's qubits
[e.g.~this could happen if Alice instructed him to apply a Hadamard
gate to an equally weighted superposition of the two qubits initially
in $|0\>$]. In this case she may disentangle $A$ from ${\cal M}$ by sending to Bob a
new instruction which {\em undoes} the previous transformation, and
then she can measure the qubits of $A$ and check whether they match
with her original decoy.  For most universal gate sets this can be
easily done without requiring nontrivial quantum processing by Alice:
when $U_n=U_n^\dag$ [e.g. for Hadamards and C-NOTs] she just needs to
send back to Bob the {\emph{same}} register $A$ she had received from
him, using it as a subsequent instruction of the computation. For the
$\pi/8$ gates the same result is obtained with $8$ consecutive
iterations. The need of bouncing back the same register multiple
times does not weaken the security of the scheme.  In fact, from
Bob's point of view this is just equivalent to him seeing such states
for a longer time: any coherent cheating strategy he can apply to the
successive iterations of such states is equivalent to a strategy that
he applies the first time he sees them.  In other words, he does not
gain any advantage from the fact that Alice is sending them multiple
times. Bob may become entangled with Alice's register {\em during} the
protocol, but, importantly, he must be disentangled before Alice
measures: he cannot retain any information by the time the computation
ends.

Summarizing, if Alice knows that the qubits she received back from Bob
are factorized from his computation qubits, she measures them and
checks whether they match the decoy state she had sent him. If,
instead, she does not know it, she bounces back to Bob the register
$A$ (either twice or $8$ times depending on the gates involved) and
then she measures it. If her measurements disagree with the
state she had originally sent him, she is certain that Bob is trying
to extract information from her queries. The protocol is summarized in
table~\ref{t:prot}.

\begin{table}\begin{center}\begin{tabular}{|l|l|}\hline
      1.&Bob initializes his qubits in $|0\>^{\otimes N}$.\\\hline
      2.&\begin{minipage}{.45\textwidth}\flushleft\vspace{2pt} $j$th computation
        step: Alice sends Bob a register $A$ of $O(\log_2 N)$ qubits.
        It (randomly)
        {\em either} contains the qubit to which a gate is to be applied
        (e.g.~$|3\>_A$ means ``apply the  Hadamard gate to qubit
        $\#3$''), {\em or} it contains a decoy
        (e.g.~$|n\>_A+|n'\>_A$). \vspace{2pt}\end{minipage}\\\hline
      3.&\begin{minipage}{.44\textwidth}\flushleft \vspace{2pt}
        Bob uses Alice's register to establish to which qubits to
        apply the gates of the universal set: 
        e.g.~Bob's action $U_{Bob}|3\>_A|\Psi\>_{\cal M}$ applies the Hadamard gate to
        Bob's qubit $\#3$ (here $|\Psi\>_{\cal M}$ represents the global
        state of Bob's qubits). If the register contains a decoy, he will apply
        the gates to a {\em superposition} of
        registers.\vspace{2pt}\end{minipage}\\\hline
      4.&\begin{minipage}{.44\textwidth}\flushleft \vspace{2pt}
        Bob sends the register $A$ back to Alice.\vspace{2pt}\end{minipage}\\\hline
      5.&\begin{minipage}{.44\textwidth}\flushleft\vspace{2pt} If
        Alice knows that the register $A$ is unentangled from Bob's
        qubits, she measures it [case (a), see text]. Otherwise she  sends it back to
        Bob as one of the  successive instructions until it becomes
        unentangled,
        and then  measures it [case (b)]. If
        the measurement result matches
        the state she had initially prepared,  she
        proceeds to the next step of the computation through
        point 2, otherwise she halts the 
        computation.\vspace{2pt}\end{minipage}\\\hline
      6.&\begin{minipage}{.44\textwidth}\flushleft\vspace{2pt}
        At the end of  the computation, Bob measures the computation
        qubits and reveals the
        computation result (possibly encrypted, see text).
        \vspace{2pt}\end{minipage}\\\hline 
\end{tabular}\end{center}
\caption{Scheme of the protocol.
  \labell{t:prot}}
\end{table}

\section{Security analysis} The protocol security is based on the fact
that Bob does not know whether each single register sent by Alice is a
decoy or a plain instruction, and these are encoded into states
belonging to non-orthogonal subspaces.
 (In this respect, it is important to notice that already
Alice's first communication can contain a decoy for all possible
gates.) 
Accordingly, any information that Bob extracts from the computational
basis will disturb the superpositions of the decoy states
\cite{fuchs,infodis}, giving Alice a nonzero probability of
identifying his cheat.
This observation can be made rigorous by exploiting a formal
connection between our scheme and
the BB84 quantum-key-distribution protocol~\cite{bb84}, whose security
is well established \cite{mayers,lochau,shor,elle} also when
significantly different probabilities are assigned to the
complementary coding bases $\{|0\>,|1\>\}$ and
$\{|+\>,|-\>\}$~\cite{loarde}.  Specifically, our protocol can be
recast as an instance of a (non-balanced~\cite{loarde}) BB84 where
Alice sends a secret key to herself using Bob as a quantum channel:
she is transmitting and recovering sequences of $|0\>$'s and $|1\>$'s
(plain instructions) and sequences of $|0\>$, $|1\>$, $|\pm\>$'s
(decoys).  Since Alice already knows the key she is sending to
herself, she can easily determine whether
Bob is disturbing the channel by trying to read the key, i.e.~by
acquiring information on the computation Alice wishes him to perform
[the actions of an honest Bob 
are carefully designed not to disturb the key so that he will not be
mistaken as a cheater].  Because of the formal mapping detailed above,
the security of the whole procedure is guaranteed by the security of
the BB84 protocol.

The probability that Bob can cheat for $j$ computational steps without
being detected by Alice decreases exponentially as $p^{\gamma j}$,
where $\gamma$ is the average fraction of instructions that are decoys
and $p$ is the probability of being detected on a single decoy. So on
average he can cheat only for a constant number of steps before
triggering Alice's cheat detection. He will thus be able to obtain
information only on a constant fraction of Alice's gates before she
stops the protocol: asymptotically in $J$ he obtains no information on
Alice's computation. In contrast to the cheat-sensitivity of the QPQ
or of the protocol of \cite{arrighi}, a cheating Bob cannot obtain the
full information here (strong cheat-sensitivity).

\section{Resource accounting}
We now give a resource-accounting of the protocol. As stated above,
this protocol is optimal in terms of the number of exchanged bits of
information between Alice and Bob, as a universal quantum computer
cannot have a software register of less than $O(J\log_2N)$ bits. The
blind protocol simply requires these to be qubits instead of bits and
requires a small overhead composed by the decoys and the final
one-time-pad encoding (see below). The total communication complexity
is then still $O(J\log_2N)$ qubits. The running time overhead will be
linear (a constant fraction $\gamma$ of the operations will be decoy
operations), so the algorithm running time will be $O((1+\gamma)J)$.
The fraction $\gamma$ can be chosen arbitrarily small for $J\to\infty$
as (ignoring logarithmic corrections) it can scale as $J^{-1/2}$, see
\cite{loarde}. The overhead in terms of qubits is linear: we need a
constant fraction $\lambda$ of qubits to be devoted to Alice's decoy
operations, so that the qubit cost goes from $N$ to $N(1+\lambda)$. In
terms of gates, there is no overhead with respect to what is necessary
for a universal programmable quantum computer \cite{nc}. The only
difference is that the software is encoded in quantum bits (Alice's
register) instead of a sequence of classical numbers. This means that
Bob needs controlled-swaps that are controlled by a quantum register
instead of a classical register that would be sufficient for a
programmable quantum computer. Summarizing, except for logarithmic or
constant corrections, the blind computation protocol proposed here
does not require a significant computational or communication overhead
over what is necessary for a universal programmable quantum computer:
the only substantial difference is that the software (Alice's
registers) is encoded in qubits instead of bits.

\section{Add-ons} 
In addition to guaranteeing the privacy of Alice's computation, our
protocol ensures also privacy of the {\em input} and {\em output}
data. The protection of the input is a trivial consequence of the fact
that the preparation of the initial state is included in the algorithm
[remember that Bob's quantum computer starts from a fiducial state
e.g.~$|0\>^{\otimes N}$].  To protect the output Alice can instruct
Bob to perform random bit flips on the computation qubits in the same
basis on which he will perform the computation's final measurement.
This means that Bob's outcome will be randomized with a one-time-pad
of which only Alice has the key: it will be secure from anyone else.

In the protocol described up to now Alice cannot ascertain whether Bob
is indeed performing the computation she has requested. Even a
non-cheating Bob could still be uncooperative and perform a different
computation. Interestingly, even though Alice has limited
computational capabilities, she can still check that Bob is
cooperating using the ideas of ``interacting proofs'' described in
\cite{abe} and adapted to blind computation in \cite{bfk,FK,NS}. The basic
idea is very simple: hidden in her computation, Alice places some trap
qubits. Since Bob does not know the position of the traps, he will
flip the trap qubits with high probability if he does not follow
Alice's instructions. It is also possible to use quantum error
correction codes to increase the probability that an uncooperative Bob
flips the trap \cite{abe,FK,NS,bfk}.

\section{Conclusions}
We presented a scheme for performing universal blind quantum
computation where both Alice's algorithm and her input-output data are
hidden from anyone else. It is efficient in terms of communication and
computational resources. It is a cheat-sensitive scheme: if Bob tries
to extract information from Alice's registers or to perform a
different computation, she can find it out and stop the protocol
before he gains a significant fraction of the information on the
computation.

We thank A. Ac\'{i}n for pointing out Ref.~\cite{loarde}. LM and TR
acknowledge support by the Royal Society for attending the
International Seminar on Sources and signatures of quantum
enhancements in technology. TM is supported by JSPS and Program to
Disseminate Tenure Tracking System by MEXT, TR by the Leverhulme
Trust.


\begin{references}
\bibitem{NOTA1} Due to Alice's computational limitation, she is not
  able to prepare some of the possible decoy states, such as the
  superposition $|n=0\>_A+|n=3\>_A$ which is an entangled state
  $|0,0\>_A+|1,1\>_A$ in binary notation.  This limitation does not
  impair the security of the protocol, since Alice can still prepare
  decoys states where the plain instruction $|n=0\>_A$ and $|n=3\>_A$
  are superimposed: e.g., the unentangled state
  $|n=0\>_A+|n=1\>_A+|n=2\>_A+|n=3\>_A$, which in binary notation
  corresponds to the factorized state~$|+,+\rangle_A$.
\end{references}
\end{document}